# Subthreshold Swing Behavior in Amorphous Indium-Gallium-Zinc-Oxide Transistors from Room to Cryogenic Temperatures

Hongwei Tang[1,2,a)], Attilio Belmonte[1)], Dennis Lin[1)], Ying Zhao[1)], Arnout Beckers[1)], Patrick Verdonck[1)], Harold Dekkers[1)], Subhali Subhechha[1)], Michiel van Setten[1)], Zhuo Chen[1,2)], Gouri Sankar Kar[1)], Jan Van Houdt[1,2)], and Valeri Afanas'ev[1,2)]

[1] *imec, Kapeldreef 75, 3001 Leuven, Belgium*

[2] *Department of Physics and Astronomy, Katholieke Universiteit Leuven, Celestijnenlaan 200d, 3001 Leuven, Belgium*

While cryogenic-temperature subthreshold swing (*SS*) in crystalline semiconductors has been widely studied, a careful study on the temperature-dependent *SS* in amorphous oxide semiconductors remains lacking. In this paper, a comprehensive analysis of the *SS* in thin-film transistors with an amorphous indium gallium zinc oxide (IGZO) channel at temperatures from 300 K down to 4 K is presented. Main observations include: 1) At room temperature (300 K), the devices exhibit a SS of 61 mV/dec, and a low interface trap density ($<10^{11}$ cm$^{-2}$), among the best reported values for IGZO devices. 2) A SS saturation around 40 mV/dec is observed between 200 K and 100 K. It is well explained by the electron transport via band tail states with exponential decay ($W_t$) of 13 meV. 3) At deep-cryogenic temperature, SS increase significantly exceeding 200 mV/dec at 4 K. Such high SS values are actually limited by the measurement current range, confirmed by $I_d$-$V_g$ simulations based on the variable range hopping (VRH) model. This work not only elucidates the *SS* behavior in amorphous IGZO devices but also provides a deep understanding of the physical mechanisms of electron transport in amorphous semiconductors.

Subthreshold swing (*SS*) is a fundamental parameter in semiconductor transistors, representing the gate voltage required to increase the drain current by one decade in the subthreshold region[1]. A lower *SS* help reduce power consumption and steeper switching characteristics. *SS* also provides crucial insights into the density of interface traps[2,3], making it a key parameter for understanding device physics. At room temperature or above, *SS* typically exhibits a linear temperature dependence as described by the Boltzmann relation ($SS \approx ln10 \times k_B T/q$, blue line in **Figure 1**)[1]. At cryogenic temperature, recent studies[4–10] have revealed significant deviations from this behavior. Below certain low temperatures, the *SS* shows no further decrease (**Figure 1**) and saturates to values in the range from 10 mV/dec to 30 mV/dec. This phenomenon has been extensively reported in various transistor technologies, such as bulk-Si FET[4], floating-body silicon-on-insulator (FD-SOI)[5], SiGe nanowire FET[10] and gallium nitride (GaN)[11]. The saturation behavior is typically explained by band tailing due to disorder (surface roughness, defects) at the semiconductor-gate dielectric interface[12], substrate impurity[7] or source to drain tunneling[13].

---

a) Corresponding author email: hongwei.tang@imec.be

In the field of amorphous oxide semiconductors, such as indium-gallium-zinc-oxide (IGZO), previous reports[14–17] have reported *SS* behavior at different temperatures, but have not identified *SS* saturation in IGZO. This raises an important and interesting question: does *SS* saturation occur in amorphous semiconductor devices? Moreover, it has been found that amorphous semiconductor devices [15,16,18,19] exhibit a unique behavior at deep-cryogenic temperatures, where *SS* increases rather than decreases - a clear contrast to the trends observed in crystalline semiconductors. An effective physical explanation for this anomaly remains elusive.

Furthermore, recent studies in silicon-based DRAM have shown that cryogenic operation can enhance retention time through decreasing *SS* [20,21]. Since IGZO is also a promising candidate for capacitor-less DRAM[22], understanding its SS behavior help assess its potential in cryogenic applications (e.g. the memory for quantum computer operating below a few tens of K[23]).

To address this, we present a detailed investigation of the temperature-dependence *SS* in IGZO transistors across a broad temperature range from 4 to 300 K. Three distinct trends are identified: a linear decrease below room temperature down to 200 K, *SS* saturation in the intermediate temperature range from 200 K to 100 K, and abnormal *SS* increase at deep cryogenic temperatures. We correlate these phenomena with the amorphous nature of the semiconductor and the electron transport mechanism different from crystalline semiconductor materials.

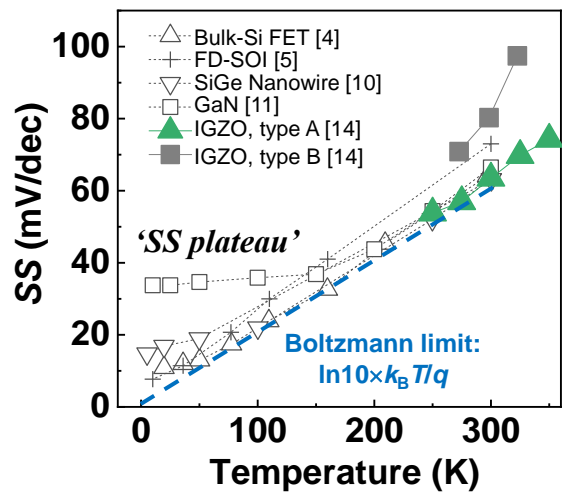

**Figure 1.** Summary of subthreshold swing (*SS*) behavior at various temperatures of different transistor technologies, showing a plateau of *SS* which deviates from the Boltzmann limit.

In this study, we utilized a back-gate (BG) transistor structure (**Figure 2a**) with an amorphous IGZO channel. The device uses a heavily p-doped silicon substrate as the back gate, with a 20 nm thermally grown $SiO_2$ layer as gate dielectric. The channel is a 10 nm amorphous IGZO layer deposited via physical vapor deposition (PVD), and then encapsulated by a 100 nm $SiO_2$ layer deposited using plasma-enhanced chemical vapor deposition (PECVD). The contact electrodes comprise of a stack of 10 nm Ti and 200 nm Au. During fabrication, the device undergoes dry air annealing twice (250°C/1 hour) to passivate oxygen-deficiency defects (oxygen vacancies) in IGZO. Electrical measurements were conducted using Agilent B1530A semiconductor parameter analyzer. Gate voltage is applied to the Si substrate. For cryogenic temperature measurements, the sample was loaded in a vacuum probe station equipped with a liquid helium cooling system. The sample was initially cooled to 4 K and subsequently measured during the heating process back to room temperature 300 K.

**Figure 2b** illustrates the $I_d$ - $V_g$ characteristics of BG IGZO device measured across the temperature range from 4 K to 300 K at $V_d = 0.05$ V. The device width ($W_{ch}$) / length ($L_{ch}$) are 1280 μm and 3 μm respectively. The threshold voltage ($V_{th}$) becomes more positive as the temperature decreases. This can be attributed to multiple factors. First, the Fermi level ($E_F$) moves closer to the conduction band edge, increasing the flat-band voltage. Second, acceptor-like traps below the $E_F$ become filled with electrons, introducing negative charges that make $V_{th}$ more positive. Additionally, the on-current ($I_{on}$) decreases significantly with decreasing temperature due to changes in charge transport mechanisms. These observations of $V_{th}$ shift and $I_{on}$ reduction align well with previous studies on amorphous semiconductor devices[15,16,24].

To explore subthreshold characteristics, the *SS* values were extracted as $\partial \log 10(I_d)/\partial V_g$ and plotted as a function of $I_d$ in **Figure 2c**. In the range of 4 K to 50 K, *SS* decreases with decreasing $I_d$ but remains higher than 60 mV/dec. Notably, some *SS* points fall below 60 mV/dec at temperatures from 100 K to 300 K, as further highlighted in **Figure 2d**. A clear trend emerges: *SS* decreases steadily as the temperature drops from 300 K to 200 K, remaining nearly constant across two orders of magnitude of current. However, between 200 K and 100 K, *SS* reaches a plateau around 40 mV/dec and does not decrease further. (Although the lowest *SS* value of approximately 33 mV/dec is recorded at $T = 125$ K, this measurement may be unreliable because the corresponding $I_d$ approaches the noise level of $10^{-13}$ A. We therefore averaged the last two data points for a more reliable estimate.)

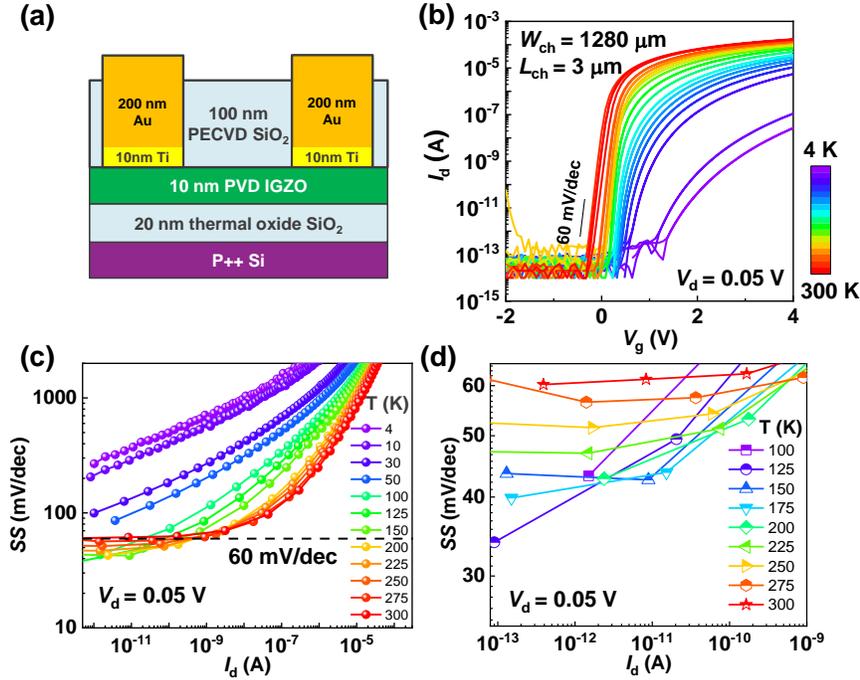

**Figure 2**. (a) Schematic illustration of the back-gated IGZO transistor. (b) $I_d$ - $V_g$ characteristics of the IGZO transistor under 4 K to 300 K. Below 50 K, the temperatures are 4 K, 10 K, 30 K and 50 K. From 50 K to 300 K, each step is 25 K. (c) The plot of $SS$ versus $I_d$ under different temperatures at $V_d$ = 0.05 V. The black dash line is the Boltzmann limit at 300 K which is 60 mV/dec. (d) The zoom-in region which shows data points below $SS$ = 60 mV/dec at temperatures lower than 300 K. Note that the color for each temperature in panel (d) is different from panels (b) and (c).

**Figure 3a** summarized the $SS$ as a function of temperatures, where $SS$ values are extracted as the minimum slopes above a reliable current level (~$10^{-12}$ A). Some key points are highlighted: $SS$ = 61 mV/dec at 300 K, 40 mV/dec at 175 K and 267 mV/dec at 4 K. A zoom-in region is shown in **Figure 3b** and **3c**. As the temperature decreases from 300 K to 200 K (Region I), $SS$ decreases linearly with temperature, consistent with the Boltzmann relation. In fully depleted IGZO thin films, $SS$ can be expressed as[3]: $SS \approx (1 + qD_t/C_{ox})\, ln10 \times k_B T/q$, where $k_B$ is Boltzmann constant, $q$ is electron charge, $D_t$ is the total trap density and $C_{ox}$ is dielectric capacitance (0.18 μF/cm$^2$). From these calculations, $D_t$ is found to be below $10^{11}$ cm$^{-2}$, which is remarkably low for amorphous IGZO devices.

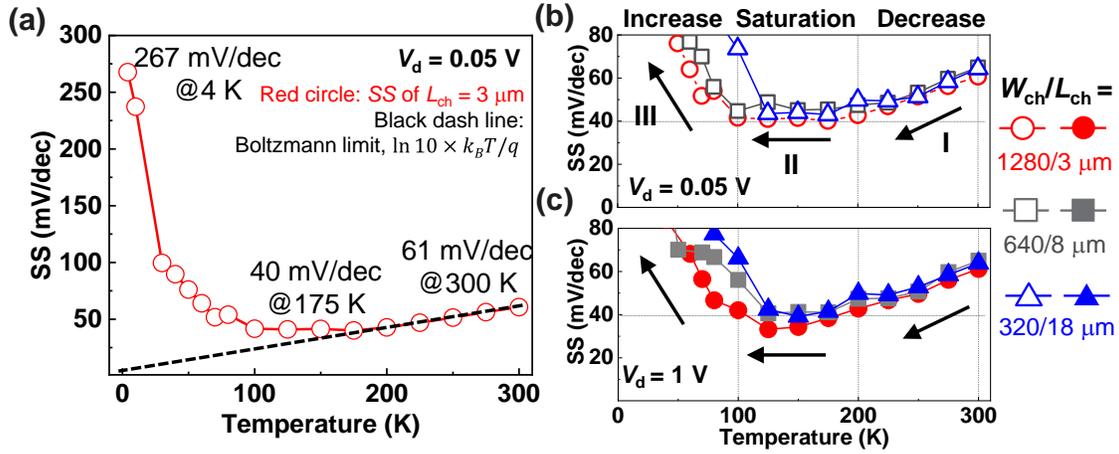

**Figure 3.** (a) Extracted *SS* across a wide temperature range, illustrating distinct trends from room to cryogenic temperatures. The *SS* values are extracted as minimum slopes above the current level (~$10^{-12}$ A) in **Figure 2b**. (b) Zoomed-in view of the low-*SS* region for devices with varying channel widths ($W_{ch}$) and lengths ($L_{ch}$) at $V_d = 0.05$ V and (c) $V_d = 1$ V. *SS* increases around 100 K and becomes higher for smaller $W_{ch}/L_{ch}$ ratios due to current degradation, a similar situation as illustrated in **Figure 5a**.

In Region II (100 - 200 K), *SS* exhibits a saturation-like behavior across different drain voltages ($V_d = 0.05$ V and 1 V) and device dimensions ($W_{ch}/L_{ch} =$ 1280 / 3, 640 / 18 and 320 / 18 μm), as shown in **Figure 3b** and **3c**. This finding confirms that *SS* saturation in IGZO devices is a reproducible phenomenon and reflects the amorphous IGZO properties, making it the first experimental observation of its kind.

Previous studies have developed theoretical models based on crystalline silicon properties to explain *SS* saturation[4,5,12,25]. For example, Beckers *et al*. proposed a *'physics-based'* expression for Si MOSFETs based on a band-tail hypothesis[12]. They argued that exponential band tails originate from Gaussian-distributed potential wells in the Si channel. Ghibaudo *et al*. further extended this derivation using the Kubo–Greenwood formalism[25]. Despite differences in their approaches, both studies conclude that below a characteristic temperature ($T_c$), *SS* follows the simple relation: $SS(T < T_c) = ln(10)k_B T_c/q$, where $T_c$ represents the characteristic temperature where *SS* starts to saturate.

The presence of band tail states provides a compelling explanation for the *SS* phenomena in IGZO devices. It is reported that statistical compositional variations of indium, gallium, and zinc components in amorphous IGZO matrix induce random band edge fluctuations (**Figure 4a**) across the channel plane[26], resulting in band tail states with density decaying exponentially with energy as $exp(\frac{E-E_c}{W_t})$, where $W_t = k_B T_c$ is the slope of the band tail, $E_c$ is the conduction band edge (**Figure 4b**). At low temperatures, Fermi-Dirac distribution function (f(E)) approaches a step function (**Figure 4c,** blue line), and the characteristic width of band tail becomes the dominant limiting factor for *SS*.

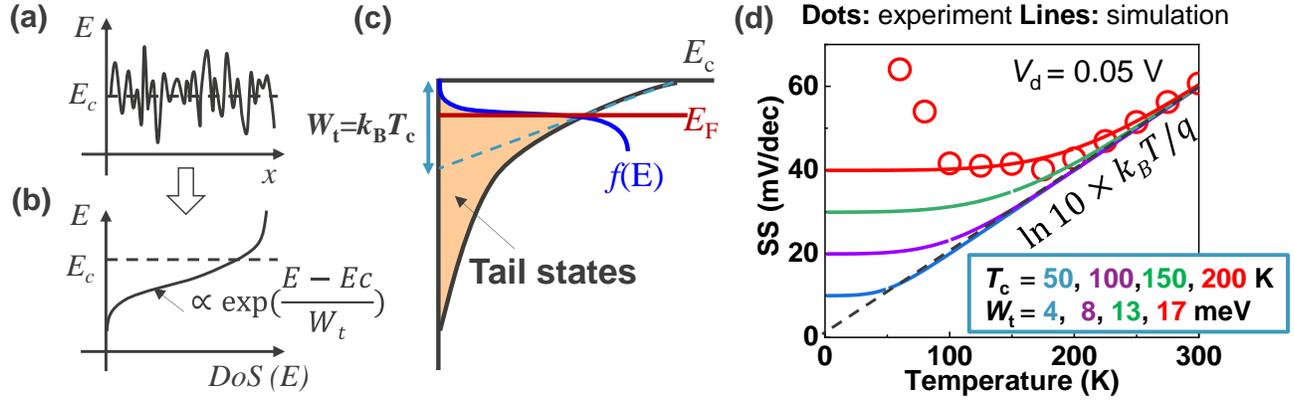

**Figure 4**. (a) Sketch of band edge fluctuation of amorphous IGZO, leading to (b) band tail states below conduction band edge ($E_c$). (c) Illustration of the integration of band tail states and Fermi-Dirac distribution function $f(E)$. $W_t$ is the width of band tail and $T_c$ is the critical temperature. (d) Comparison of the simulated *SS* plateau (lines) for different $W_t$ and $T_c$ values with experimental data (dots), demonstrating a strong match for $W_t$ = 17 meV.

Based on the derivation of the *SS*-temperature dependence[12], we simulated SS under different $T_c$ / $W_t$ and compared the results with experimental data (**Figure 4d**). In our experiment, the onset of SS saturation appears around 200 K with a value of 40 mV/dec. The trend closely matches the simulation plateau at $T_c$ = 200 K, confirming the validity of this model. The extracted $W_t$ is approximately 17 meV, which is also consistent with typical values reported for IGZO devices[27,28].

At temperatures below 100 K, *SS* begins to increase (Region III in **Figure 3b**). Although some studies suggest[15,19] that higher trap densities near the $E_c$ contribute to this increase, such an explanation would imply an unreasonably high $D_t$. As demonstrated in Region II, higher trap densities lead to *SS* saturation rather than an increase. Therefore, the increase in SS at cryogenic temperatures is likely due to additional mechanisms.

Here we consider that at deep cryogenic temperatures, electron transport in amorphous IGZO is dominated by variable-range hopping (VRH) conduction[15,17,18,29]. The current-temperature relationship in VRH conduction can be expressed as[29]: $I \propto \exp\left[-\left(\frac{T_0}{T}\right)^{\frac{1}{4}}\right]$, where $T_0$ is a characteristic temperature related to the density of states (DoS) and localization length. This relationship predicts the current to decrease exponentially with decreasing temperature.

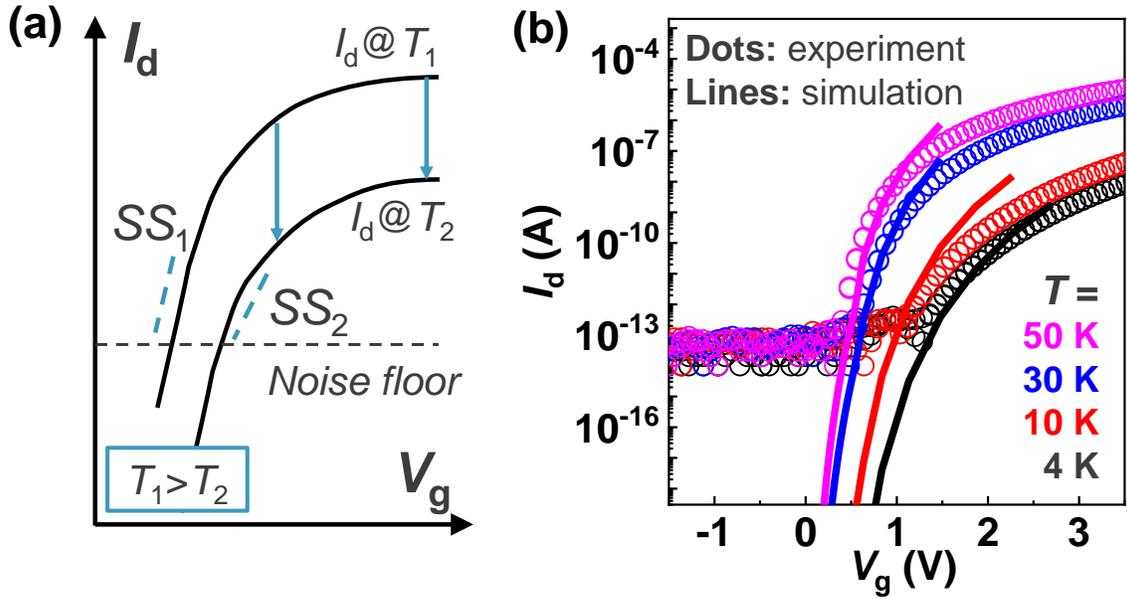

**Figure 5**. (a) Schematic illustration of current degradation as temperature decreases, leading to a higher *SS* above current noise floor. (b) Simulated data using VRH and experimental $I_d$ - $V_g$ data of 4, 10, 30 and 50 K. Solid lines are simulated data while solid dots are experimental data.

A simple illustration can explain how current degradation can lead to an increase in *SS* (**Figure 5a**). As the temperature drops from $T_1$ to $T_2$, the current decreases by several orders of magnitude, resulting in a lower $I_d$ at $T_2$. Because *SS* is defined as the minimum slope of the current curve above the detection limit, the degraded current at $T_2$ forces the inferred *SS* ($SS_2$) to be higher than that at $T_1$ ($SS_1$).

The underlying physics suggests that, to achieve a given current level, a higher electron concentration is required at lower temperatures. Since IGZO devices work in accumulation mode, the surface potential need to shift from deep depletion towards the transition region to weak accumulation. As a result, the ratio of gate voltage to surface potential increases, leading to higher *SS* values. This mechanism also explains why *SS* curves are not flat even at very low $I_d$ (**Figure 2c**) under $T$ = 4 - 50 K. Notably, this behavior is not observed in crystalline silicon devices, since higher mobility and current at cryogenic temperatures[30] prevents such degradation.

Based on the VRH model, we simulated the $I_d$ - $V_g$ behavior (**Figure 5b**) from 4 K to 50 K. The simulation successfully captures the general trend, including the observed increase in *SS* with decreasing temperature. It also clearly shows that at lower current levels, *SS* becomes steeper even at deep cryogenic temperatures (e.g. *SS* < 100 mV/dec at 4 K) - a regime that is challenging to be measured directly. In future work, indirect extraction methods using 2T0C/2T1C cell structures[31,32] could offer a approach to accurately extract ultra-low off-currents and boost the understanding of *SS* behavior at cryogenic temperatures.

Some mismatches at $T = 4$ K and 10 K may arise from effects like carrier freeze-out[33] and trap-assisted tunneling[34], which have not been included in our model. Although they affect fitting precision, both support the observed current degradation at lower temperatures, which is the main reason to explain the high *SS* in the measurable $I_d$ range.

In conclusion, we presented a comprehensive analysis of the temperature-dependence of *SS* of amorphous n-channel IGZO field-effect transistors. The fabricated devices exhibit an excellent *SS* of approximately 61 mV/dec at 300 K which saturates around 40 mV/dec at 200 K. The *SS* saturation is explained by the existence of the conduction band tail states, extending the understanding from crystalline materials to amorphous semiconductors. Below 100 K, the *SS* increases significantly, exceeding 200 mV/dec at 4 K. This abnormal increase is because the extraction of the minimum *SS* is limited by the current detection range. The $I_d$-$V_g$ simulation at cryogenic temperatures further supports this explanation. We believe this work offers a deeper understanding of IGZO transistor performance across different temperatures and provide insights into device physics of amorphous semiconductors.

The authors would like to thank the support of imec's Industrial Partners in the Active Memory Program. This work has been enabled in part by the NanoIC pilot line. The acquisition and operation are jointly funded by the Chips Joint Undertaking, through the European Union's Digital Europe (101183266) and Horizon Europe programs (101183277), as well as by the participating states Belgium (Flanders), France, Germany, Finland, Ireland and Romania. For more information, visit nanoic-project.eu.